\documentclass[floatfix,twocolumn,nofootinbib,superscriptaddress,a4paper,preprintnumbers]{revtex4-1}

\usepackage{amsmath}
\usepackage{amssymb}
\usepackage{graphicx}
\usepackage{ulem}
\usepackage[mathscr]{eucal}
\usepackage{color}
\usepackage[T1]{fontenc} 
\usepackage{slashed}
\usepackage{xspace}
\usepackage{hyperref}
\usepackage{xspace}
\usepackage{listings}
\usepackage{gensymb}

\newcommand{\MS}{$\overline{\text{MS}}$\xspace}

\newcommand\SARAH{{\tt SARAH}\xspace}
\newcommand\SPheno{{\tt SPheno}\xspace}

 \definecolor{mkgreen}{rgb}{0.2,.70,.3}

\begin{document}

\preprint{KA-TP-12-2018}
\preprint{BONN-TH-2018-04}


\title{Unitarity constraints in triplet extensions beyond the large $s$ limit}


\author{Manuel E. Krauss}
\email{mkrauss@th.physik.uni-bonn.de}
\affiliation{Bethe Center for Theoretical Physics \& Physikalisches Institut der Universit\"at Bonn, Nu{\ss}allee 12, 53115 Bonn, Germany}

\author{Florian Staub}
\email{florian.staub@kit.edu}
\affiliation{Institute for Theoretical Physics (ITP), Karlsruhe Institute of Technology, Engesserstra{\ss}e 7, D-76128 Karlsruhe, Germany}
\affiliation{Institute for Nuclear Physics (IKP), Karlsruhe Institute of Technology, Hermann-von-Helmholtz-Platz 1, D-76344 Eggenstein-Leopoldshafen, Germany}



\begin{abstract}
Triplet extensions are attractive alternatives to the standard model (SM) of particle physics. While models with only one triplet are highly constrained by electroweak precision observables, this is not necessarily the case once
several triplets are present as in the Georgi-Machacek model. As in all other BSM models, the parameter space of triplet extensions is constrained by the condition that perturbative unitarity is not violated. For this purpose, limits on the eigenvalues of the scalar $2\to 2$  scattering matrix are set.  It is very common in the BSM literature that the scattering matrix is calculated under one crucial assumption: the scattering 
energy $s$ is so large that only point interactions involving quartic couplings provide non-negligible contributions. 
However, it is not given that this approximation is always valid -- in fact, diagrams involving propagators can play an important role. We discuss at the examples of (i) the SM model extended by a real triplet, (ii) the $Y=1$ triplet extension of the SM, and (iii) the Georgi-Machacek model, how the tree-level unitarity constraints are affected once the large $s$ approximation is given up. For all models we find that the impact of (effective) cubic couplings can be crucial. 
\end{abstract}
\maketitle

\section{Introduction}
While the LHC continues ruling out more and more parameter space of models beyond the SM (BSM), a central question remains unanswered: whether the measured Higgs boson is the only electroweak- to TeV-scale scalar particle, or if there is more in the scalar sector which takes part in electroweak symmetry breaking (EWSB). Many attractive models have been proposed which are either motivated from theory as they solve or ameliorate problems in the standard model, or from experimental reasons in the sense that they provide new interesting signals which could be measured at colliders.
One of these possibilities is the presence of one or more triplet scalars. Those could either be introduced for implementing a seesaw mechanism of type~II for generating small Majorana neutrino masses \cite{Schechter:1980gr,Cheng:1980qt}, or for providing an alternative to EWSB where triplets actively contribute by developing non-negligible vacuum expectation values (VEVs) like in the so-called Georgi-Machacek (GM) model  \cite{Georgi:1985nv}.

Before analysing its properties for the LHC, each model's parameter space has to be confronted with theoretical constraints. 
Among the most stringent are the conditions that perturbative unitarity of scalar $2\to 2$ scattering must not be violated. For the simplest triplet extension, for instance, those have been derived in Ref.~\cite{Khan:2016sxm} and for the GM model in
Refs.~\cite{Aoki:2007ah,Hartling:2014zca}. 
These derivations, as almost all other unitarity constraints on BSM models which are 
found and applied in the literature, make use of the limit that the scattering energy $s$ is much larger than all involved masses. This has the benefit that all diagrams containing propagators can be disregarded and only quartic point interactions need to be taken into account. 

 Recently, it has been pointed out in Ref.~\cite{Krauss:2017xpj} that also checks of the perturbative behaviour of a non-supersymmetric model should be taken seriously. It was shown at the example of the GM model how large loop corrections can be triggered by large scalar trilinear couplings, ultimately casting doubt about the perturbative treatment in large regions of parameter space.
These large couplings might not be visible at first glance when trading the corresponding parameters for the (tree-level) masses and therefore removing them from the list of ``input'' parameters. 

However, large trilinear couplings do not only affect the loop corrections in a model, but also the $2 \to 2$ scattering processes at $\sqrt{s}$ not much larger than the involved masses. As a result, the amplitude at finite $\sqrt{s}$ might be significantly larger than in the typical limit $\sqrt{s} \to \infty$, therefore also affecting the perturbative unitarity constraints!
Because of that, we check here the full scattering matrix for all possible $2\to 2$ processes with external scalars.
We do so for a simple real triplet extension, a complex triplet extension, and ultimately the GM model.
 We do not only include the possibility of finite $\sqrt{s}$ but also the effects of electroweak symmetry breaking. While this kind of calculation was done for the SM decades ago \cite{Lee:1977eg}, the impact of trilinear couplings on the unitarity constraints in BSM models were to our knowledge only checked for the minimal supersymmetric SM \cite{Schuessler:2007av} and 
 singlet extensions \cite{Cynolter:2004cq,Kang:2013zba,DiLuzio:2016sur}.\footnote{In Ref.~\cite{DiLuzio:2017tfn}, the effect was included for obtaining bounds on the Higgs trilinear coupling.}
We show how the unitarity constraints at finite energies cut deeply into the otherwise allowed parameter space of all 
considered example models and compare them, for the GM model, with the loop-improved unitarity and perturbativity checks discussed in Ref.~\cite{Krauss:2017xpj}.
 
This paper is organised as follows: in sec.~\ref{sec:unitarity} we show the main ingredients for the calculation of the perturbative unitarity checks. In secs.~\ref{sec:real-triplet}, \ref{sec:complex-triplet}
 and \ref{sec:GM}, we present the three example models and show the resulting additional constraints coming from the 
 inclusion of the improved treatment of the scalar scattering amplitudes. We conclude in sec.~\ref{sec:conclusion}.

\section{Tree-Level perturbative unitarity constraints}
\label{sec:unitarity}
\subsection{Approximations vs. full calculation}
Perturbative unitarity constraints consider the 
 $2 \to 2$ scalar field scattering amplitudes. This means that the 0th partial wave amplitude $a_0$ must satisfy either $|a_0| \leq 1$ or $|{\mathcal Re} [ a_0]| \leq \frac{1}{2}$.  
The matrix $a_0$ is given by

\begin{equation}
a_0^{ba} = \frac{1}{32\pi} \sqrt{\frac{4 |\vec{p}^{\,b}| |\vec{p}^{\,a}|}{2^{\delta_{12}} 2^{\delta_{34}}\, s}} \int_{-1}^1 d(\cos \theta) \mathcal{M}_{ba} (\cos \theta)\,,
\end{equation}
where $\vec p^{\,a(b)}$ is the center-of-mass three-momentum of the incoming (outgoing) particle pair $a=\{1,2\}~(b=\{3,4\})$, $\theta$ is the angle between these three-momenta and $\mathcal M_{ba}(\cos\theta)$ is the scattering matrix element. The exponents $\delta_{ij}$ are $1$ if the  particles $i$ and $j$ are identical, and zero otherwise.

At the tree level, the $2 \to 2$ amplitudes are real, which is why one usually uses the more severe constraint  $|{\mathcal Re} [ a_0]| \leq \frac{1}{2}$, which leads to $|\mathcal{M}| < 8 \pi$ in the limit $s\to \infty$. 
This must be satisfied by all of the eigenvalues $\tilde{x}_i$ of the scattering matrix $\mathcal{M}$. $\mathcal{M}$ must be derived by  including each possible combination of two scalar fields in the initial and final states.
 
For analysing whether perturbative unitarity is given or not, it is common to work in the high energy limit, i.e. the dominant tree-level diagrams contributing to $|\mathcal M|$ involve only quartic interactions. All other diagrams with propagators are suppressed by the collision energy squared and are neglected. 
 Moreover, effects of electroweak symmetry breaking (EWSB) are usually ignored, i.e. Goldstone bosons are considered as physical fields. 
 
However, it is hardly tested if the large $s$ approximation is valid in all BSM models in which it is applied. It could be that large contributions are present at small $s$ which then rule out given parameter regions in the considered model. 
Just consider for instance a large TeV-scale cubic scalar interaction $\kappa \phi_i\phi_j\phi_k$. Above the resonance, a typical diagram would therefore scale with $\kappa^2/s$ and hence be relevant for $\sqrt{s}$ of $\mathcal O({\rm TeV})$.

In order to be able to apply these tests, the {\tt Mathematica} package \SARAH has now been extended by the functionality to derive more reliable unitarity limits by giving up the large $s$ approximations. Details of this implementation in \SARAH are given elsewhere~\cite{Goodsell:2018tti}. We only want to summarise the main aspects:
\begin{itemize}
 \item All tree-level diagrams with internal and external scalars are included to calculate the full scattering matrix 
 \item The calculation is done in terms of mass eigenstates, i.e. the full VEV dependence is kept
 \item All necessary routines for a numerical evaluation with \SPheno are generated 
 \item Very large enhancements close to poles or kinematic thresholds are cut in order not to overestimate the limits
 \item Renormalization group equation (RGE) running can be included to obtain an estimate of the higher order corrections
\end{itemize}

\subsection{Analysis Setup}
We are going to study the impact of the improved unitarity constraints on three triplet extensions of the SM: (i) with one real triplet, (ii) with one complex triplet, as well as (iii) with both one complex and one real triplet, the Georgi-Machacek model. Our numerical analysis will be based on the \SPheno~\cite{Porod:2003um,Porod:2011nf} interface of \SARAH~\cite{Staub:2008uz,Staub:2009bi,Staub:2010jh,Staub:2012pb,Staub:2013tta}.  By default, {\tt SPheno} calculates the mass spectrum  at the full one-loop level  and includes all important two-loop corrections to the neutral scalar masses \cite{Goodsell:2014bna,Goodsell:2015ira,Braathen:2017izn}. However, we are not making use of these routines in the following but work under the assumption that an on-shell (OS) calculation is working in principle (with all the caveats discussed in Ref.~\cite{Krauss:2017xpj}). Thus,  only the tree-level masses are calculated. These are then used to calculate the perturbative unitarity constraints for a given scattering energy $\sqrt{s}$. The constraints from Higgs searches are included via {\tt HiggsBounds} \cite{Bechtle:2008jh,Bechtle:2011sb,Bechtle:2013wla}.

\section{The real triplet extended Standard Model}
\label{sec:real-triplet}
\subsection{Model description}
We start with a rather simple BSM model: the SM extended by a real scalar $SU(2)$-triplet $T$ without hypercharge
\begin{equation}
T = \begin{pmatrix}
     T^0/\sqrt{2} & T^- \\ (T^-)^* & - T^0/\sqrt{2}
    \end{pmatrix}
\end{equation}
The scalar potential of this model is given by
\begin{align}
V = & m_H^2 |H|^2 + \frac12 m_T^2 \text{Tr}(T^2) + \frac12 \lambda_H |H|^4  + \frac12 \lambda_T \text{Tr}(T^4) \nonumber \\
 & \hspace{2cm} + \frac12 \lambda_{HT} |H|^2 \text{Tr}(T^2) + \kappa H^\dagger T H
\end{align}
After EWSB, both the Higgs as well as the neutral component of the triplet receive a vacuum expectation 
value:
\begin{equation}
\langle T^0 \rangle = \frac{1}{\sqrt{2}} v_T \,, \hspace{1cm} \langle H^0 \rangle = \frac{1}{\sqrt{2}} v\,.
\end{equation}

The scalar mass eigenstates are two CP even states which are a mixture of $H^0$ and $T^0$ with masses $m_h$ and $m_H$, as well as a (physical) charged 
Higgs boson $H^\pm$ with mass $m_{H^+}$ which is  a mixture of $H^+$ and $T^-$. The rotation in the neutral Higgs sector is fixed by an angle $\alpha$. 
Therefore, it is possible to trade the four Lagrangian parameters $\lambda_i$ ($i=H,T,HT$) and $\kappa$ for the three scalar masses and one rotation angle. 
The relations are:
\begin{align}
\kappa & = \frac{2 m_{H^+}^2 v_T}{\tilde v^2} \,,\\
 \lambda_H & = \frac{m_h^2+m_H^2 t_\alpha^2}{\left(t_\alpha^2+1\right) v^2} \,,\\
 \lambda_{HT} & = \frac{1}{v v_T}\left(\frac{\sqrt{2} t_\alpha (m_h^2-m_H^2)}{t_\alpha^2+1}+\frac{2 m_{H^+}^2 v v_T}{\tilde v^2}\right) \,,\\
 \lambda_{T} & = \frac{\tilde v^2 m_H^2-m_{H^+}^2 \left(t_\alpha^2+1\right) v^2+m_h^2 t_\alpha^2 \tilde v^2}{\left(t_\alpha^2+1\right) v_T^2 \tilde v^2} \,,
\end{align}
where we have defined $\tilde v = \sqrt{2 v_T^2 + v^2}$.
Since $v_T$ must be small in order not to be in conflict with electroweak precision data \cite{Gunion:1990dt,Arhrib:2011uy,Kanemura:2012rs,Maiezza:2016bzp}, $m_H$ and $m_{H^+}$ must always be close in order to avoid too large quartic couplings. In addition, $\kappa$ needs to be small.
The absolute values of the  eigenvalues of the scattering matrix
in the limit of large $s$ are given by: 
\begin{eqnarray*}
&8 \pi > \text{Max} \Big\{|\lambda_H|, |\lambda_{HT}|, 2 |\lambda_{T}|,& \\
&\frac{1}{2} \left| -3 \lambda_H-5 \lambda_{T}\pm \sqrt{9 \lambda_H^2-30 \lambda_{T} \lambda_H+12 \lambda_{HT}^2+25 \lambda_{T}^2}\right|\Big\}\,.&
\end{eqnarray*}

\subsection{Results}
\begin{figure}
\includegraphics[width=0.7\linewidth]{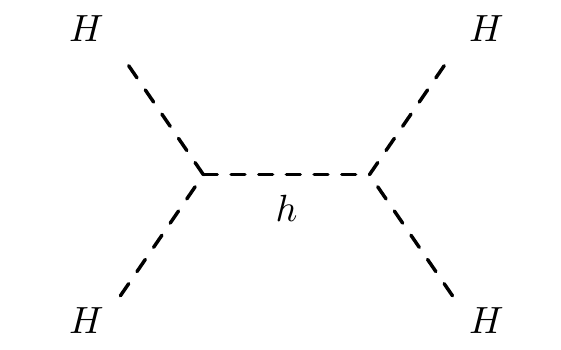} 
\caption{Diagram contributing to the scalar scattering matrix as finite $s$.}
\label{fig:diagram_real}
\end{figure}

\begin{figure}[tb]
\includegraphics[width=\linewidth]{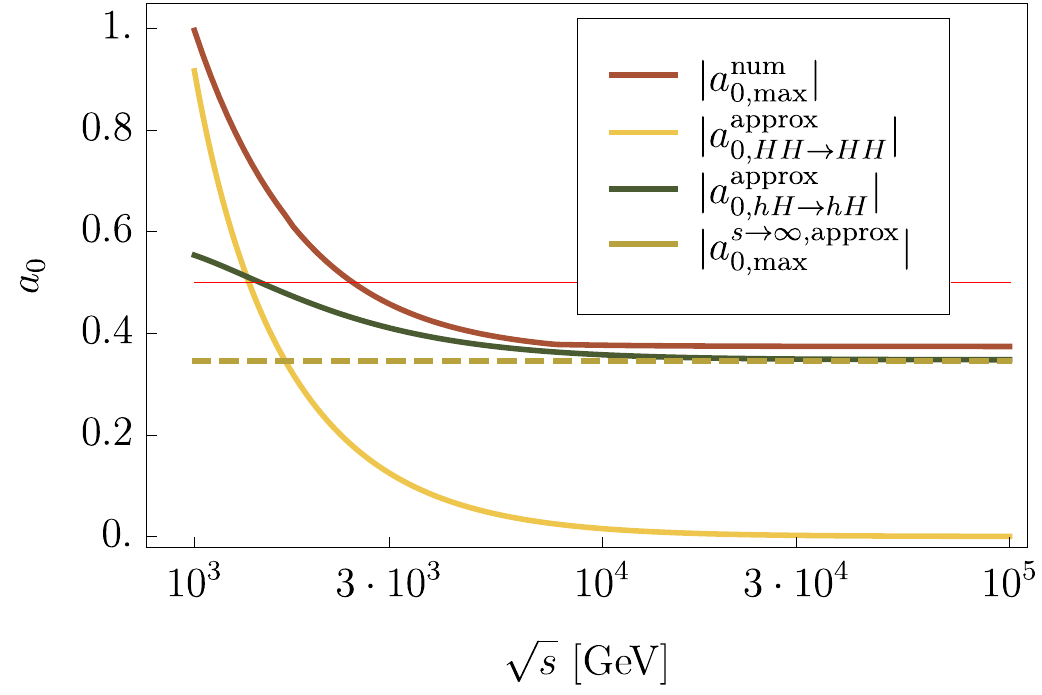}
\caption{Comparison of the approximated values for the $HH\to HH$ and $hH\to hH$ scatterings at finite $\sqrt{s}$ with the full numerical calculation and the approximated result for $s\to\infty$. 
We have used here $m_H = 400~\text{GeV},\tan\alpha=0.07, v_T=3~\text{GeV}$, and $m_{H^+}$ was fixed by the condition $\lambda_T=0$.}
\label{fig:tsm_s}
\end{figure}

In order to show the importance of additional tree-level contributions to the scattering matrix as a function of $\tan\alpha$ and $m_H$, we force $\lambda_T$ to be small. This can be done by setting $m_{H^+}$ to
\begin{equation}
m_{H^+} = \frac{\sqrt{m_H^2 + m_h^2 t_\alpha^2} \, \tilde v}{\sqrt{1 + t_\alpha^2 }\, v}\,.
\end{equation}
Thus, $\lambda_T$ vanishes and the largest eigenvalue is approximately given by
\begin{equation}
a_0^{s\to\infty}\simeq \frac{3 m_h^2+\frac{m_H^2 \left(3 v_T t_\alpha^2+2 \sqrt{6} \sqrt{t_\alpha^2 v^2-2 \sqrt{2} t_\alpha v_T v+2 v_T^2}\right)}{v_T}}{32 \pi    v^2} 
\end{equation}
Here, we assumed $\tan\alpha \ll 1 $ and $m_H \gg m_h$. We learn from this that unitarity tends to be violated for increasing values of $m_H$ and $\tan\alpha$  and decreasing 
values of $v_T$. \\
The large $s$ approximation needs to be compared for instance with the diagram shown in Fig.~\ref{fig:diagram_real}.
The $s$-channel diagrams is of the form
\begin{equation}
\frac{|c|^2}{m_h^2-s} 
\end{equation}
where the vertex $c$ in the limit $\lambda_{HT} \gg \lambda_T,\lambda_H$ is given by
\begin{eqnarray}
c=&&\frac{1}{8} \big(\sqrt{2} (\sin\alpha-3 \sin(3 \alpha )) (\kappa+\lambda_{HT} v_T)+2 \lambda_{HT} v \cos (\alpha ) \nonumber \\
&&\hspace{1.cm} +6 \lambda_{HT} v \cos (3 \alpha )\big) \,,
\end{eqnarray}
which for small $\tan\alpha$ can be further  simplified to
\begin{equation}
c \simeq -\sqrt{2} t_\alpha (\kappa + \lambda_{HT} v_T) +  \lambda_{HT} v \simeq    \lambda_{HT} v\,.
\end{equation}
Thus, for $\sqrt{s}$ not much larger than $m_h$ one can expect that this diagram scales as $\frac{\lambda_{HT}^2 v^2}{m_h^2}$. Therefore, although the cubic Lagrangian parameter is small, the EWSB-generated terms lead to sizeable contributions by diagrams of the type of Fig.~\ref{fig:diagram_real}.
Actually, a more careful calculation including also the crossed 
$t$- and $u$-channel diagrams results in
\begin{eqnarray}
a_0&&(HH\to HH) \simeq \nonumber \\
-&&\frac{m_H^4 \left(t_\alpha v - \sqrt{2}v_T\right)^2 }{16 \pi  v^2 v_T^2 \sqrt{s^3 \left(s-4 m_H^2\right)}} \times \nonumber \\&& \times \left(2 s \log \left(\frac{m_h^2}{m_h^2-4 m_H^2+s}\right)-4 m_H^2+s\right) \,.\nonumber \\
\end{eqnarray}
For somewhat larger $s$ the dominant contribution to the full scattering matrix comes from the process $hH \to hH$ which can be approximated to
\begin{eqnarray}
&& a_0(hH\to hH) \simeq \nonumber \\
&& \frac{m_H^2}{16 \pi  s v^2 v_T^2 \left(m_H^2-s\right)} \Big(\left(s-m_H^2\right)t_\alpha v \big[m_H^2 \left(2 t_\alpha v - 3 \sqrt{2} v_T \right)  \nonumber \\ 
&&  -\sqrt{2} s v_T\big] +2 m_H^2 s \log  \left(\frac{m_H^2}{s}\right) \left(t_\alpha v - \sqrt{2}v_T\right)^2\Big)\,. \nonumber \\
\end{eqnarray}
Of course, one needs to keep in mind that these are just single entries in the scattering matrix which needs to be diagonalised. However, we will not quote here any analytical 
approximations when doing that since they are not very helpful because of their length. We compare in Fig.~\ref{fig:tsm_s} the analytical approximations for $a_0(hH\to hH)$, 
$a_0(HH\to HH)$ and $a_0^{s\to\infty}$ with the full numerical calculation. As an example we have chosen here 
\begin{equation}
m_H = 400~\text{GeV},\tan\alpha=0.07, v_T=3~\text{GeV} \,.
\end{equation}
The range of $\sqrt{s}$ is chosen such that all possible resonances are avoided -- and starts about an order of magnitude above the $s$-channel resonance of Fig.~\ref{fig:diagram_real} which leads to the largest contribution at small scattering energies.
We can see that for small $s$ the full numerical result agrees quite well with the approximation for the $s$-channel $HH\to HH$ scattering and is significantly larger than the limit $a_0 < \frac12$ -- i.e., unitarity is violated!
 For larger $\sqrt{s}$, the full numerical result approaches this approximate asymptotic value, i.e. the diagrams including trilinear interactions become suppressed.
If one uses only the approximation of large $s$, it would seem that $a_0 < \frac12$ is fulfilled, therefore underestimating the actual constraints.
In the end, this particular combination of model parameters is forbidden since it violates perturbative unitarity -- which is only seen by including the effects of EWSB-generated trilinear scalar interactions at finite scattering energies.

\begin{figure}[tb]
\includegraphics[width=\linewidth]{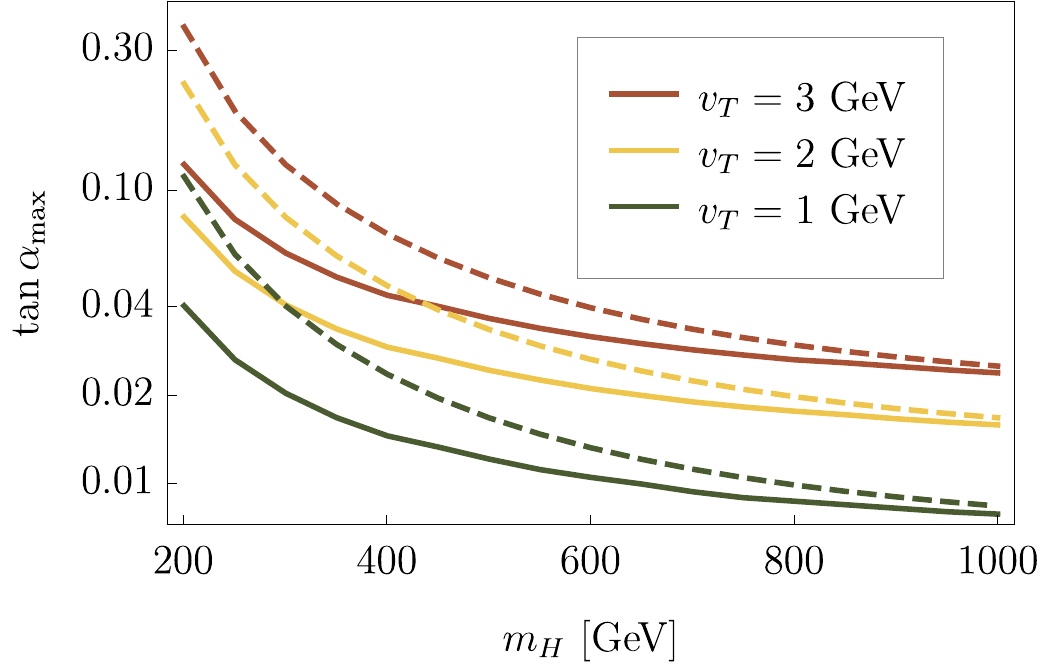}\\
\includegraphics[width=\linewidth]{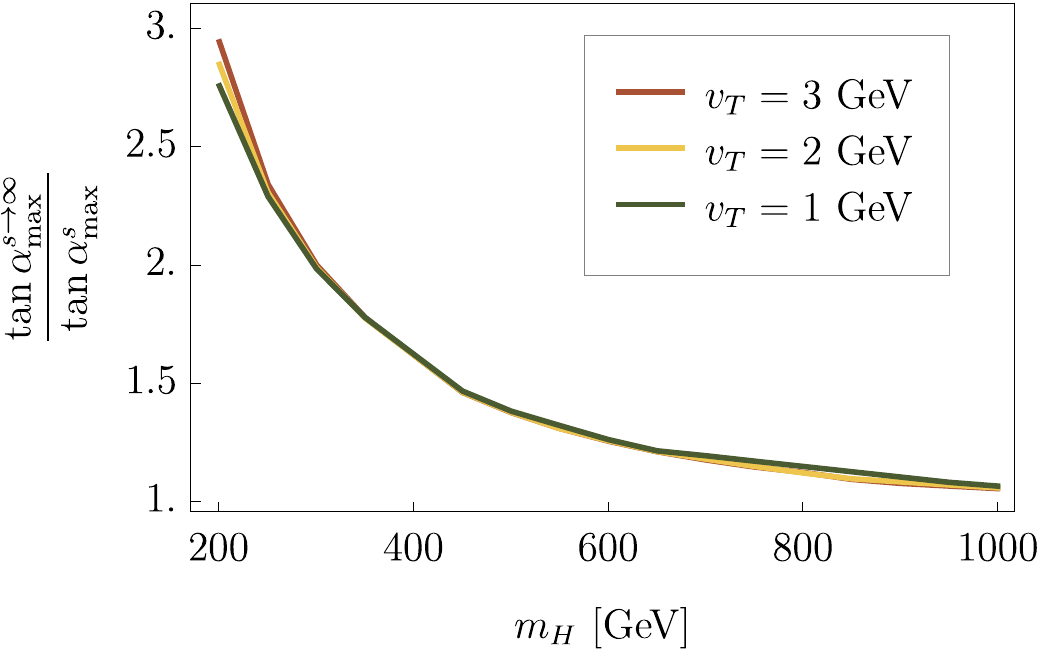}
\caption{First row: Comparison between the old (dashed lines) and new (full lines) unitarity constraints for the SM extended by a real triplet. Here, 
we show the maximally allowed value of $\tan\alpha_{\rm max}$ for a given heavy Higgs mass $m_H$ and three different values of the triplet VEV $v_T$. $m_{H^+}$ is chosen to obtain $\lambda_T=0$. In the second 
row the ratio of  $\tan\alpha_{\rm max}$ for the full calculation and the large $s$ approximation is shown.}
\label{fig:tsm_mh_ta}
\end{figure}

We now check how the difference between the full tree-level calculation and the large $s$ approximation is affected by the different parameters. For this purpose, we show in Fig.~\ref{fig:tsm_mh_ta} the maximal allowed value of $\tan\alpha$ for given values of $m_H$ and the triplet VEV $v_T$. We find as expected that the maximal value of $\tan\alpha$ quickly drops for larger $m_H$ and smaller $v_T$ for both calculations, the one with and without  explicit $s$ dependence. However, we also see that for smaller $m_H$ a large difference exists between both calculations: the value allowed for $\tan\alpha$ based on the point interactions only is about a factor of 3 larger than the correct one by including all contributions. This difference becomes smaller for increasing $m_H$. The ratio for $\tan\alpha_{\rm max}$ for both calculations is nearly independent of the chosen value of $v_T$ as can be seen in the second row of Fig.~\ref{fig:tsm_mh_ta}.

Finally, we want to remark that we were only concerned with the improved unitarity constraints in this model.
 In addition to our findings, a recent study found that also the impact of the modified Veltman conditions can place very severe limits on this model, see Ref.~\cite{Chabab:2018ert} for details.

\section{The complex Triplet extension of the SM}
\label{sec:complex-triplet}
\subsection{Model description}

The general scalar potential for the SM extended by a complex $SU(2)_L$ triplet with hypercharge $Y=1$ 
can be written as\footnote{Here and in the following we use the convention $Q_{\rm em} = T_{3L}+Y$.}
\begin{align}
V &=   m_H^2 |H|^2 + m_T^2 \text{Tr}(T^\dagger T) 
+ \left(
\kappa H^T  T^\dagger H 
+ \text{h.c}\right)
   \nonumber \\
& + \frac12 \lambda_H |H|^4 + \frac12 \lambda_T \text{Tr}(T^\dagger T T^\dagger T)
  -  \frac12 \lambda_{HT} |H|^2 \text{Tr}(T T^\dagger) \nonumber \\
& +  \frac12 \lambda_{HT'} H^\dagger T^\dagger T H +  \frac12 \lambda_{T'} \text{Tr}(T^\dagger T)^2  \,.
\end{align}
The matrix form for the triplet is given by
\begin{align}
T
= & \begin{pmatrix} T^+/\sqrt{2} & - T^{++} \\ T^0 & - T^+/\sqrt{2} \end{pmatrix} \,.
\end{align}
The scalar spectrum in this model consists of two neutral CP even,  one neutral CP odd, two charged and one doubly-charged Higgs with masses $m_h, m_H, m_A, m_{H^+}, 
m_{H^{++}}$. The relations between the masses and the Lagrangian parameters are
\begin{align}
\lambda_H &=  \frac{m_h^2+m_H^2 t_\alpha^2}{\left(t_\alpha^2+1\right)} v^2 \,,\\
\lambda_{HT}&= -\frac{4 m_A^2}{v^2+4 v_T^2}-\frac{2 t_\alpha (m_h-m_H) (m_h+m_H)}{\left(t_\alpha^2+1\right) v v_T} \,,\\
\kappa&= \frac{\sqrt{2} m_A^2 v_T}{v^2+4 v_T^2} \,,\\
\lambda_{T'}&=   \frac{m_A^2 v^2}{v^2 v_T^2+4 v_T^4}+\frac{8 m_{H^+}^2}{v^2+2 v_T^2}  \\
 & +\frac{m_h^2 t_\alpha^2+m_H^2 -(4 m_{H^+}^2-2 m_{H^{++}}^2)\left(t_\alpha^2+1\right)}{\left(t_\alpha^2+1\right) v_T^2} \nonumber \,,\\
\lambda_{HT'} &=  \frac{8 m_{H^+}^2}{v^2+2 v_T^2}-\frac{8  m_A^2}{v^2+4 v_T^2}\,,\\
\lambda_{T}&= -\frac{2 \left(m_A^2-2 m_{H^+}^2+m_{H^{++}}^2\right)}{v_T^2} \nonumber \\
 & +\frac{8 m_A^2}{v^2+4 v_T^2}-\frac{8  m_{H^+}^2}{v^2+2 v_T^2}\,,
\end{align}
The mass terms $m_H^2$ and $m_T^2$ are fixed by the minimization conditions of the scalar potential.

\subsection{Results}
\begin{figure}[tb]
\includegraphics[width=\linewidth]{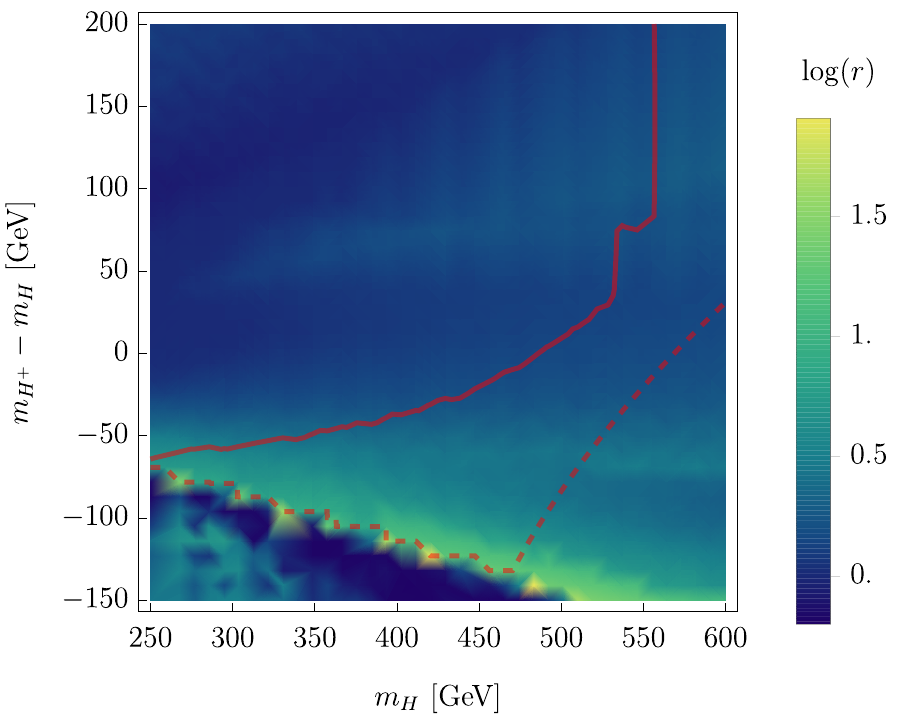}\\
\includegraphics[width=\linewidth]{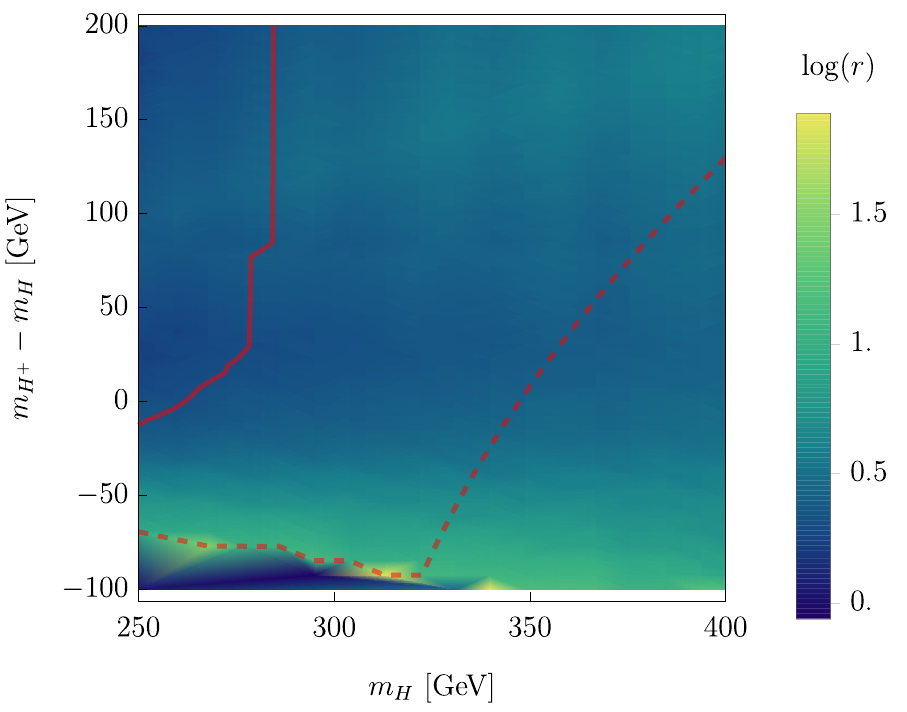}
\caption{Exclusion limits using the unitarity constraints in the large $s$ limit (dashed line) and the full calculation (full lines) in the $(m_H,m_{H^+}-m_H)$ plane. 
Parameter space to the right of the contour lines is excluded by the respective conditions. 
The colour code shows the logarithm of the ratio of the maximal eigenvalues of the scattering matrix for the two calculations. In the first row, we use $\tan\alpha=0.05$, and in the second $\tan\alpha=0.1$. The masses $m_A$ and $m_{H^{++}}$ are fixed by the conditions $\lambda_T=\lambda_{T'}=0$.}
\label{fig:yTSM}
\end{figure}

The unitarity constraints in the large $s$ limit are given by
\begin{align}\label{eq:y=1_sLargeLimit}
& \text{Max}\Big\{
\left| \lambda_H\right|, 
\frac{1}{2} \left| \lambda_{HT}\right|, \frac{1}{4} \left| 2 \lambda_{HT}-3 \lambda_{HT'}\right|, 
\frac{1}{2} \left| \lambda_{HT}-\lambda_{HT'}\right|, \nonumber \\
& \frac{1}{4} \left| 2 \lambda_{HT}+\lambda_{HT'}\right| , 
\frac{1}{2} \left| \lambda_{T}-2 \lambda_{T'}\right| , 
\left| \lambda_{T'}\right|,  \left| \lambda_{T}+\lambda_{T'}\right|,  \nonumber \\ 
&\frac{1}{4} \big| \pm \sqrt{4 (-3 \lambda_H+3 \lambda_T+4 \lambda_{T'})^2+6 (\lambda_{HT'}-2 \lambda_{HT})^2} \nonumber \\
 & \hspace{1cm} +6 \lambda_H+6 \lambda_T+8 \lambda_{T'}\big|, \nonumber \\
&\frac{1}{2} \left| \pm \sqrt{(-\lambda_H+2 \lambda_T+\lambda_{T'})^2+\lambda_{HT'}^2}+\lambda_H+2 \lambda_T+\lambda_{T'}\right|
\Big\} \nonumber \\
 & \hspace{2cm} < 8 \pi\,.
\end{align}
We see that in this version it is also not possible to have a large cubic interaction which is not proportional to a quartic coupling, because $\kappa$ is always smaller than $\lambda_{HT} v$. 
Nevertheless, there is an important difference compared to the case of the real triplet. While it was not possible in the real case to have a sizeable mass splitting between the additional neutral and charged 
Higgs states, this is no longer the case here. If we fix $m_A$ and $m_{H^{++}}$ by the conditions $\lambda_T=\lambda_{T'}=0$, and assume $\lambda_H$ to be small compared to $\lambda_{HT}$, $\lambda_{HT'}$, the condition to preserve unitarity becomes 
\begin{equation}
\frac{1}{2} \sqrt{\frac{3}{2}} \left| \frac{4 m_H^2 t_\alpha v-8 m_{H^+}^2 v_T}{v^2 v_T}\right| < 8 \pi\,.
\end{equation}
Thus, $m_H$ and $m_{H^+}$ do not have to be close in mass to fulfil this condition. In contrast, $m_{H^+} > m_H$ is even preferred for $t_\alpha v > v_T$.   On the other hand, the process $hh\to HH$ is in this limit calculated to
\begin{align}
&a_0(hh\to HH) = \frac{m_H^2 (t_\alpha v-2 v_T)}{32 \pi  \sqrt{s} v^2 v_T^2 \sqrt[4]{s \left(s-4
   m_H^2\right)}} \times\nonumber \\
   &\Big[2 m_H^2 (2
   v_T-t_\alpha v) \log \left(\frac{-2 m_h^2-\sqrt{s \left(s-4
   m_H^2\right)}+s}{-2 m_h^2+\sqrt{s \left(s-4
   m_H^2\right)}+s}\right)\nonumber \\
   & \hspace{2cm} +v_T \sqrt{s \left(s-4
   m_H^2\right)}\Big]\,,
\end{align}
which is completely independent of $m_{H^+}$ and grows quickly with $m_H$. As long as we have a scattering energy $\sqrt{s} < 4 m_{H^+}$, also no interferences can occur, i.e. we can expect a different behaviour compared to the large $s$ limit. This is confirmed in Fig.~\ref{fig:yTSM}, where we find a hard limit for large $m_H$ which is nearly independent of $m_{H^+}$ if the charged Higgs is heavier than the neutral one. While the constraints according to Eq.~(\ref{eq:y=1_sLargeLimit}) rule out part of the considered parameter space, the full numerical results including the trilinear interactions constrain the model more severely.
Here, the scattering energies which place the strongest constraints range from $\sim 400\,$GeV to $\sim 1\,$TeV.

\section{The Georgi-Machacek Model}
\label{sec:GM}
\label{sec:model}

\subsection{Model description}
The GM model extends the SM by two scalar $SU(2)_L$ triplets, one real triplet $\eta$ with hypercharge $Y=0$ and a complex one $\chi$ with $Y=-1$, therefore also featuring a doubly-charged component. Both triplets are required to actively participate in EWSB and  therefore to assume a VEV $v_{\eta,\phi}$. In  order not to be in conflict with electroweak precision measurements, and in particular the $\rho$ parameter, a 
global $SU(2)_L \times SU(2)_R$ symmetry is imposed, leading to $v_\eta = v_\phi$ and a simplification of the 
scalar potential. In order to write the latter in a very compact way, it is convenient to express 
the two triplets as a bi-triplet and the Higgs doublet $\Phi$ as a bi-doublet under $SU(2)_L \times SU(2)_R$:
\begin{align}
\Phi =
\left(
\begin{array}{cc}
\phi^{0*}  & \phi^+     \\
 \phi^- & \phi^0      
\end{array}
\right)\,, \quad
\quad \Delta =
\left(
\begin{array}{ccc}
 \chi^{0*} & \eta^+  & \chi^{++}   \\
 \chi^- & \eta^0  & \chi^+  \\
 \chi^{--} & \eta^{-}  & \chi^0   
\end{array}
\right)\,.
\end{align}

The scalar potential can then be written as
\begin{align}
& V(\Phi, \Delta)  =  \frac{\mu_2^2}{2}  \mathrm{Tr} \Phi^\dagger \Phi + \frac{\mu_3^2}{2} \mathrm{Tr} \Delta^\dagger \Delta  + \lambda_1 \left[  \mathrm{Tr} \Phi^\dagger \Phi \right]^2 + \nonumber \\
& \lambda_2  \mathrm{Tr} \Phi^\dagger \Phi \, \mathrm{Tr} \Delta^\dagger \Delta   + \lambda_3 \mathrm{Tr} \Delta^\dagger \Delta \Delta^\dagger \Delta +   \lambda_4 \left[\mathrm{Tr} \Delta^\dagger \Delta \right]^2   \nonumber \\
&- \lambda_5 \mathrm{Tr} \left( \Phi^\dagger \sigma^a \Phi \sigma^b  \right) \, \mathrm{Tr}  \left(\Delta^\dagger t^a \Delta t^b \right) \nonumber \\
&   - M_1  \mathrm{Tr} \left( \Phi^\dagger \tau^a \Phi \tau^b \right) (U \Delta U^\dagger)_{ab}\nonumber     \\
& - M_2 \mathrm{Tr} \left( \Delta^\dagger t^a \Delta t^b \right) (U \Delta U^\dagger)_{ab} \, ,
\end{align}
with $\tau^a$ and $t^a$ being the $SU(2)$ generators for the doublet and triplet representations; $U$ is e.g. given  in Ref.~\cite{Hartling:2014zca}. 
The electroweak VEV $v_{\rm SM}$ can then be re-expressed as 
\begin{equation}
v_{\rm SM}^2 = v_\phi^2 + 8 v_\chi^2 \simeq 246\,{\rm GeV}\,,
\end{equation}
while the relative sizes of the VEVs are parametrised by an angle $\theta_H$:
\begin{equation}
s_H \equiv \sin \theta_H = 2 \sqrt{2} \, \frac{v_\chi}{v_{\rm SM}}\,,\qquad c_H \equiv \cos \theta_H = \frac{v_\phi}{v_{\rm SM}}\,.
\end{equation}

The scalar mass spectrum of the model consists of seven physical states: three CP-even neutral scalars, one physical CP-odd scalar (i.e. pseudo-scalar), two (complex) physical singly-charged scalars and one (complex) doubly-charged Higgs. Due to the custodial symmetry, one CP-even scalar, one singly- and the doubly-charged Higgs can be combined into a custodial fiveplet with a common tree-level mass $m_5$ while the other charged scalar combines with the pseudo-scalar to a triplet with tree-level mass $m_3$.
Those masses are given by
\begin{eqnarray}
	m_5^2 &=& v_{\rm SM}\Big( s_H (3 \sqrt{2} M_2 + s_H \lambda_3 v_{\rm SM}) \nonumber \\ &+& c_H^2 (\frac{M_1}{\sqrt{2} s_H} + \frac{3}{2} \lambda_5 v_{\rm SM}) \Big)\,, \label{eq:mass_fiveplet} \\
	m_3^2 &=& 
	\frac{v_{\rm SM} M_1}{\sqrt{2}\, s_H} + \frac{1}{2} \lambda_5 v_{\rm SM}^2\,. \label{eq:mass_triplet}
\end{eqnarray}
The remaining two neutral Higgs eigenstates are denoted $h$ and $H$, where the former corresponds to the measured SM-like eigenstate. The mixing angle which rotates between these two mass eigenstates is denoted $\alpha$.
For more details on the model we refer the reader e.g. to Ref.~\cite{Hartling:2014zca}. \\\

\subsection{Unitarity constraints}
\begin{figure}
\includegraphics[width=1\linewidth]{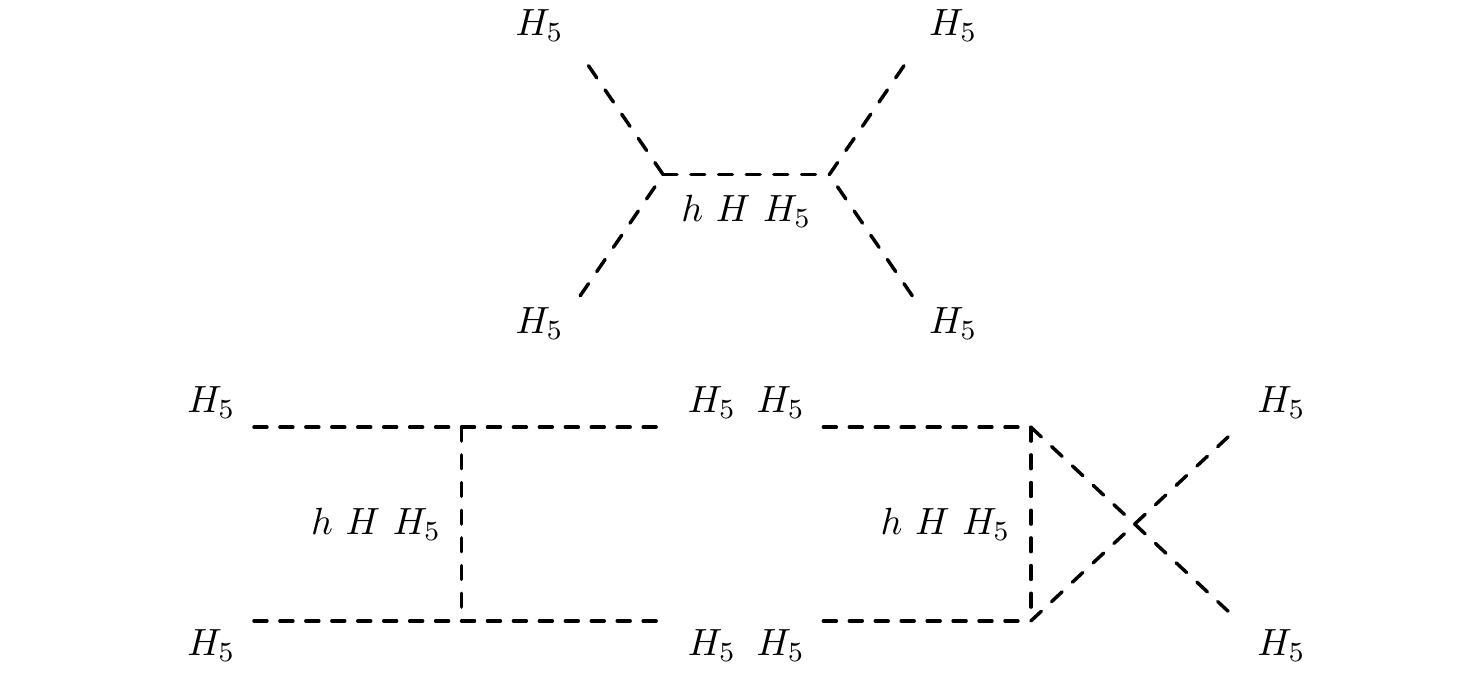} 
\caption{Diagram contributing to the scalar scattering matrix at finite $s$.}
\label{fig:H5scatter}
\end{figure}

The unitarity constraints in the large $s$ approximation for this model are given by \cite{Aoki:2007ah,Hartling:2014zca}:
\begin{eqnarray}
\sqrt{ \left( 6 \lambda_1 - 7 \lambda_3 - 11 \lambda_4 \right)^2 + 36 \lambda_2^2}
+ \big| 6 \lambda_1 + 7 \lambda_3 &+ & 11 \lambda_4   \big| \nonumber \\ &<& 4 \pi \,, \nonumber \\
\sqrt{ \left( 2 \lambda_1 + \lambda_3 - 2 \lambda_4 \right)^2 + \lambda_5^2}
+ \left| 2 \lambda_1 - \lambda_3 + 2 \lambda_4 \right| &<& 4 \pi \,, \nonumber \\
\left| 2 \lambda_3 + \lambda_4 \right| &<& \pi \,, \nonumber \\
\left| \lambda_2 - \lambda_5 \right| &<& 2 \pi \,.
\label{eq:uni}
\end{eqnarray}
In the following, we perform a brief analytical estimate of the additional terms appearing for small $s$ which can be of similar size -- or even larger. The full $2\times 2$ scattering matrix has dimension $91 \times 91$, i.e. 
it is highly unlikely to learn anything from this matrix. Therefore, we concentrate on single scattering channels like the ones depicted in Fig.~\ref{fig:H5scatter}.
We find for $\lambda_i \to 0$ ($i=2,3,4,5$) that the sum of all diagrams is given by
\begin{align}
& a_0^s(HH\to HH) =  -\frac{\sqrt{s \left(s-4 m_5^2\right)}}{216 \pi  s
   s_H^4 \left(s_H^2-1\right) \left(t_\alpha^2+1\right)^2 v_{\rm SM}^4} \times \nonumber \\
& \Bigg[m_h^2 t_\alpha v_{\rm SM} \left(-3 \sqrt{2-2 s_H^2} s_H t_\alpha+2 \sqrt{3} s_H^2-2 \sqrt{3}\right) \nonumber  \\
& + m_H^2 v_{\rm SM} \left(-2 \sqrt{3} s_H^2 t_\alpha-3 \sqrt{2-2 s_H^2} s_H+2 \sqrt{3} t_\alpha\right)\Bigg]^2 \times \nonumber \\
& \Bigg[\frac{t_\alpha^2}{t_\alpha^2+1} \left(\frac{2 \log \left(\frac{m_h^2}{-4
   m_5^2+m_h^2+s}\right)}{4 m_5^2-s}+\frac{1}{m_h^2-s}\right)+\frac{2}{m_5^2-s} + \nonumber \\
& \frac{2 \log \left(\frac{m_H^2}{-4 m_5^2+m_H^2+s}\right)}{\left(t_\alpha^2+1\right) \left(4
   m_5^2-s\right)}-\frac{4 \log \left(\frac{m_5^2}{s-3 m_5^2}\right)}{s-4 m_5^2} \nonumber \\ 
   &+\frac{1}{\left(t_\alpha^2+1\right) \left(m_H^2-s\right)}\Bigg]
\end{align}
Here, the remaining Lagrangian parameters appearing in the vertices have been re-expressed in terms of masses and mixing angles. For the relations see e.g. Ref.~\cite{Krauss:2017xpj}.
This can be further simplified if we assume $s_H,t_\alpha \ll 1$ and $m_5 \simeq 2 m_H$
\begin{align}\label{eq:GM_HH_HH}
a_0^s &(HH\to HH) = \nonumber \\ &-\frac{2 \log \left(\frac{m_H^2}{s-15 m_H^2}\right) \left(m_h^2 t_\alpha+m_H^2 (s_H-t_\alpha)\right)^2}{9 \pi  s_H^4 \left(t_\alpha^2+1\right)^3 v_{\rm SM}^2 \sqrt{s
   \left(s-16 m_H^2\right)}} 
\end{align}
Under the same assumptions, the maximal eigenvalue of Eq.~(\ref{eq:uni}) can be expressed as
\begin{equation}\label{eq:GM_HH_HH_s_inf}
a_0^{{\rm max},s\to \infty}=\frac{3 \left(m_h^2+m_H^2 t_\alpha^2\right)}{16 \pi  \left(t_\alpha^2+1\right) v_{\rm SM}^2}
\end{equation}
Thus, if we go close to the kinematic threshold, but still off-resonance, and choose $s=5\, m_5^2 \simeq 20 \,m_H^2$, we can expect that the ratio of the two expressions very roughly is
\begin{align}
\frac{a_0^s(HH\to HH)}{a_0^{{\rm max},s\to \infty}} = \frac{8 m_H^2 \log (5)}{27 \sqrt{5} s_H^2 \left(m_h^2+m_H^2 t_\alpha^2\right)} 
\end{align}
This ratio can become huge for small $s_H$ and/or large $m_H$. What is the origin of this behaviour? The vertex $c_{H_5 H_5 h}$ involved in the SM-like Higgs exchange is given by
\begin{equation}
c_{H_5 H_5 h} = -\frac{2 \sqrt{3} M_2}{\sqrt{\frac{1}{t_\alpha^2}+1}} \simeq \frac{2 m_H^2 \left(3 \sqrt{2} s_H-2 \sqrt{3} t_\alpha\right)}{3 s_H^2 \sqrt{\frac{3}{t_\alpha^2}+3} v_{\rm SM}}
\end{equation}
Thus, the huge scattering amplitudes are a consequence of large trilinear couplings which are triggered by large values of $M_2$ (and $M_1$). Since $M_2$ and $M_1$ do not enter the unitarity constraints 
so far used in the literature, this effect has been missed entirely.

\subsection{Results}
\label{sec:numerics}
In order to validate our rough analytical understanding and to further explore the impact of the new unitarity constraints, we now use the numerical machinery available with the recent 
update of \SARAH. The implementation of the GM model in \SARAH was discussed in Ref.~\cite{Staub:2016dxq}. There are many possibilities for what to use as input parameters. Naively using the Lagrangian 
parameters $\lambda_i$, $M_1$ and $M_2$ will hardly produce points which are in agreement with the Higgs mass measurements. Therefore, we trade $\lambda_1$, $M_1$ and $M_2$ for $m_h$, $m_H$ and $\alpha$. With that choice, the full set of input parameters is
\begin{eqnarray}
& m_h, m_H, \alpha, \lambda_2, \lambda_3, \lambda_4,\lambda_5, \sin\theta_H\,.
\end{eqnarray}


%

\subsubsection{Dependence on the scattering energy}
\begin{figure}[tb]
\includegraphics[width=\linewidth]{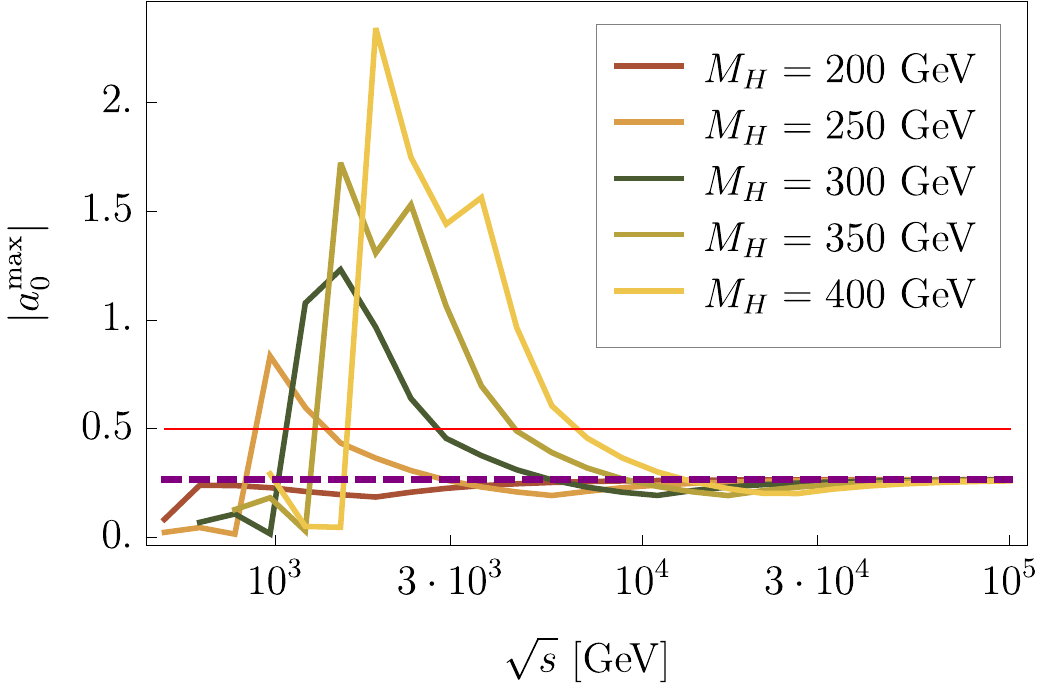}
\caption{The maximal scattering eigenvalue as a function of the scattering energy $\sqrt{s}$ for different values of $M_H$. The other parameters were set to $\lambda_2 = 0.1, \lambda_3=0.5,\lambda_4=-0.02,\lambda_5=0.1,\alpha=20\degree, \sin\theta_H = 0.15$. The dashed 
purple line gives the results using the old constraints.}
\label{fig:s-dependence}
\end{figure}
As first step, we show in Fig.~\ref{fig:s-dependence} the dependence of the full scattering matrix on the scattering energy $\sqrt{s}$ for different choices of $M_H$. The other parameters are set to
\begin{eqnarray}
&\lambda_2 = 0.1, \lambda_3=0.5,\lambda_4=-0.02,\lambda_5=0.1, & \nonumber \\
& \alpha=20\degree, \sin\theta_H = 0.15 \,.&
\label{eq:input1}
\end{eqnarray}
We can see that for $\sqrt{s}$ in the TeV range, the unitarity limits are clearly violated for $m_H$ above 250~GeV. The value of $\sqrt{s}$ at which this happens is shifted with increasing $m_H$, but at most 2.5~TeV for 
$m_H=400$~GeV. In contrast, agreement with the large $s$ approximation is only found for much larger values of $\sqrt{s}$. The difference between our full calculation and the old approximated one in the maximal scattering element from Eq.~(\ref{eq:GM_HH_HH_s_inf}) can be as large as a factor of 10 for $m_H=400$~GeV. Even for $m_H=250$~GeV a factor of 3 difference is visible.

\subsubsection{Comparison between old and new unitarity constraints}
\begin{figure}[tb]
\includegraphics[width=\linewidth]{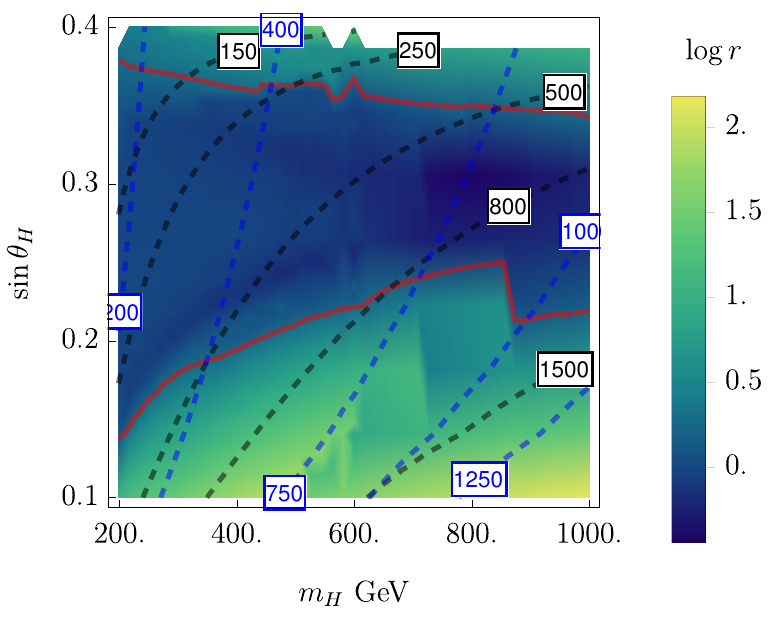} \\
\includegraphics[width=\linewidth]{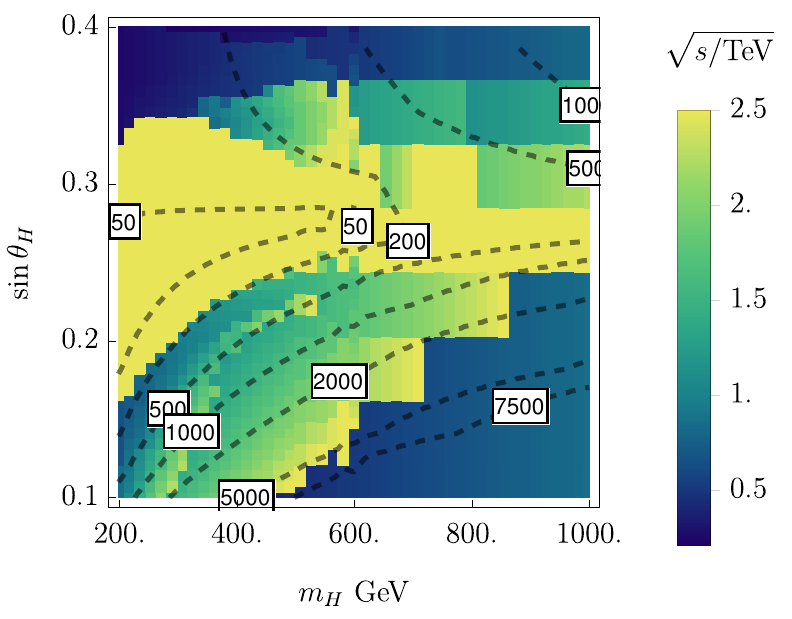}
\caption{First row: Logarithm of the ratio of the new and old results. The red line indicates the region ruled out by the new constraints, while the old calculation would allow the entire plane. The dashed contours give the values of $m_3$ (blue) and $m_5$ (black). The other parameters were set to $\lambda_2=0.1, \lambda_2=0.5, \lambda_4=-0.02, \lambda_5=0.1, \alpha=20\degree$. Here we scanned $\sqrt{s}$ between 250~GeV and 2500~TeV in order to obtain the tightest constraint. This `optimal' scattering energy is shown in the second row. The black contours in this plot show the values of $|M_2|$ in GeV.
}
\label{fig:new_vs_old}
\end{figure}
As next step, we want to make a more exhaustive comparison between the old and new results. For this purpose, we consider the $(\sin\theta_H, m_H)$ plane for the same values of $\alpha$ and $\lambda_i$ as in Eq.~(\ref{eq:input1}). In the full calculation including the propagator diagrams, we scan the scattering energy $\sqrt{s}$ between 250 and 2500~GeV to find the maximal eigenvalue of the scattering matrix. We compare this value with the one obtained by using the old constraints which only depend on the quartic couplings. The outcome is summarised in Fig.~\ref{fig:new_vs_old}. While the old constraints are passed in the entire plane, the improved calculation cuts out significant regions. This is not only the case for small $\sin\theta_H < 0.15$ as one could expect from the previous discussion, but also for $\sin\theta_H>0.35$. The reason is that $m_5$ in this parameter region scales as 
\begin{align}
m_5 & \sim m_H \sqrt{\frac{(-2 s_H \sqrt{1 - s_H^2} + \sqrt6 t_\alpha - \sqrt6 s_H^2 t_\alpha)}{s_H \sqrt{1 - s_H^2} (1 + t_\alpha^2)}} \\
& \sim 3 m_H \sqrt{\frac{-6 s_H + \sqrt{6 - 6 s_H^2}}{10 s_H}}
\end{align}
where we have used $t_\alpha \simeq \frac13$. This equation turns to zero for $s_H = 1/\sqrt{7}\simeq 0.38$, leading to tachyonic five-plets for larger $s_H$ and very small $m_5$ for $s_H$ values slightly below this root.
As consequence, the $t$-channel diagrams with a five-plet exchange become large. 
In general, we find that the maximal eigenvalue of the scattering matrix can change even by a factor of 100 for large $M_H$ and very small $\sin\theta_H$. The reason is that in this region the calculated value of $M_2$ becomes huge and is in the multi-TeV range. It is worth mentioning that this region still has a stable vacuum, i.e. the new unitarity conditions really make the difference between `allowed' or `forbidden'\footnote{This statement  holds until one also includes the perturbativity constraints which we discuss in the next subsection.}. In the second row of Fig.~\ref{fig:new_vs_old} we also show the value for the `optimal'  scattering energy, i.e. the energy at which the largest scattering eigenvalue becomes maximal, excluding resonances. In the regions which are affected most by the new constraints, this energy is moderately small and well below our largest chosen value of 2.5~TeV. 

\subsubsection{Comparison with extended perturbativity constraints}
In Ref.~\cite{Krauss:2017xpj}, a set of conditions was proposed which indicate if problems with the expansion of the perturbative series might exist. Those constraints check the relative and absolute size 
of the counter-terms (CTs) when imposing an on-shell calculation of the scalar sector.\footnote{Since the custodial symmetry is broken at the loop level, this demands to 
calculate 17 independent CTs.} A third condition makes use of the one-loop corrected couplings in the unitarity constraints on the quartic couplings after the breaking of
the custodial symmetry. The fourth condition checks the finite corrections to the scalar masses if a \MS scheme is applied instead of an OS one. We will compare the new 
unitarity constraints with these constraints, namely:
\begin{enumerate}
 \item A parameter point is considered problematic if the CT to at least one Lagrangian parameter is larger than the tree-level value of this parameter times some constant value $v$, i.e. 
 \begin{equation}
  \left|\frac{\delta x}{x}\right| > v\,.
  \label{eq:relative_size_CT}
 \end{equation}
 \item A parameter point is considered to violate perturbativity if the CT of at least one quartic coupling becomes larger than some fixed value, i.e.  
 \begin{equation}
  |\delta x| >  c \cdot \pi\,,
 \end{equation}
with  $c$ within 1 and 4. 
 \item A parameter point  is considered to violate perturbation theory  if the unitarity constraints on the quartic couplings (i.e. in the limit $s\to \infty$) are violated when inserting the renormalised couplings, i.e.  
 \begin{equation}
 |\mathcal{M}(\lambda_{Nx} \to \lambda_N + \delta\lambda_{Nx})| > 8 \pi \,. 
 \label{eq:generalized_unitarity_bound} 
 \end{equation}
\item A parameter point is considered to violate perturbation theory if the two-loop corrections to at least one scalar mass, calculated in the \MS scheme, are larger than the one-loop corrections, i.e.
 \begin{equation} 
  |(m_{\phi}^{2})_{\rm Tree} - (m_\phi^{2})_{\rm 1L}| < |(m_\phi^{2})_{\rm 2L} - (m_\phi^{2})_{ \rm 1L}| \,.
  \label{eq:size_2loop_vs_1loop}
 \end{equation}
\end{enumerate}
\begin{figure*}[tb]
\centering
\includegraphics[width=.45\linewidth]{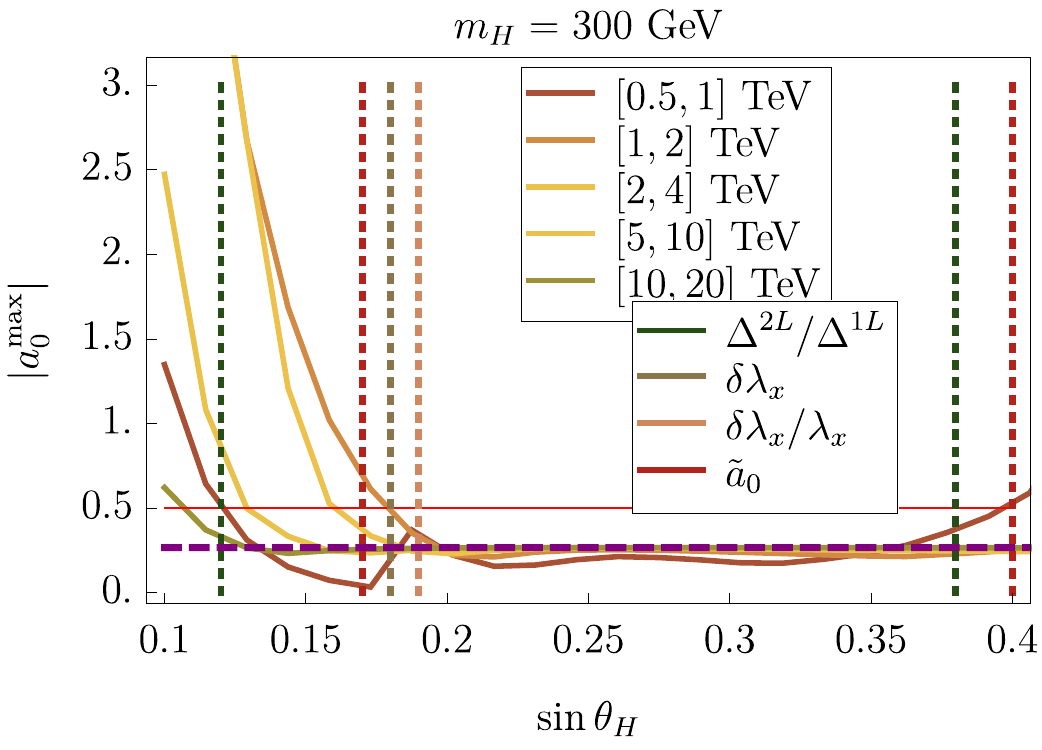} \hspace{.05\linewidth}
\includegraphics[width=.45\linewidth]{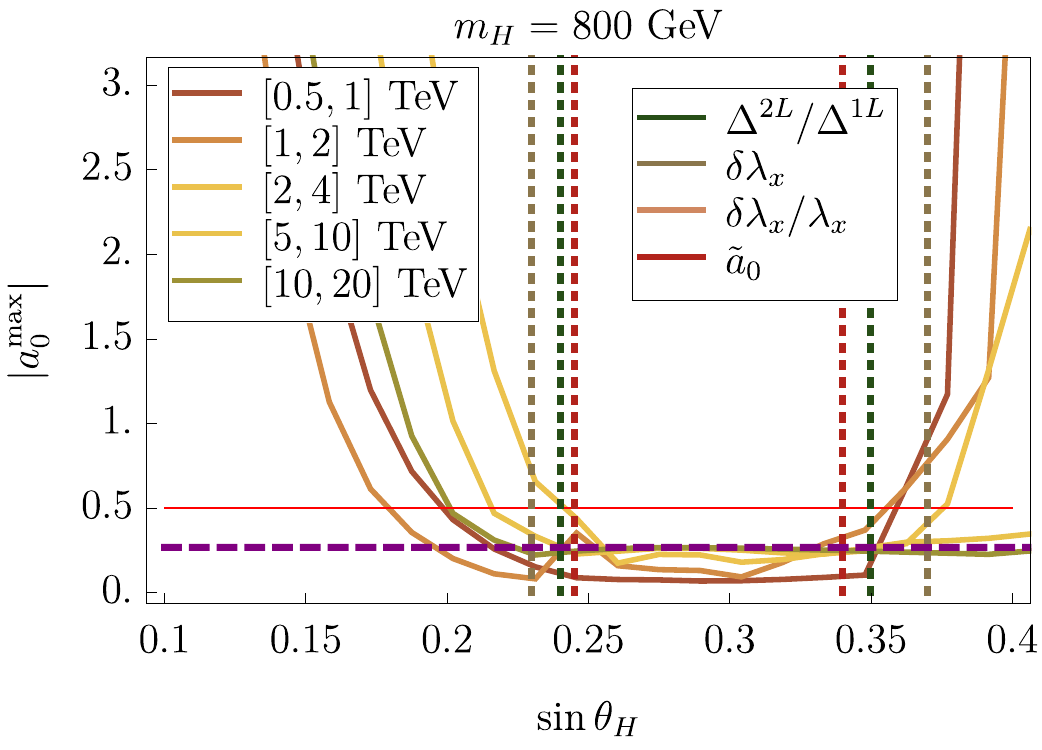} \\ \vspace{.01\textheight}
\includegraphics[width=.45\linewidth]{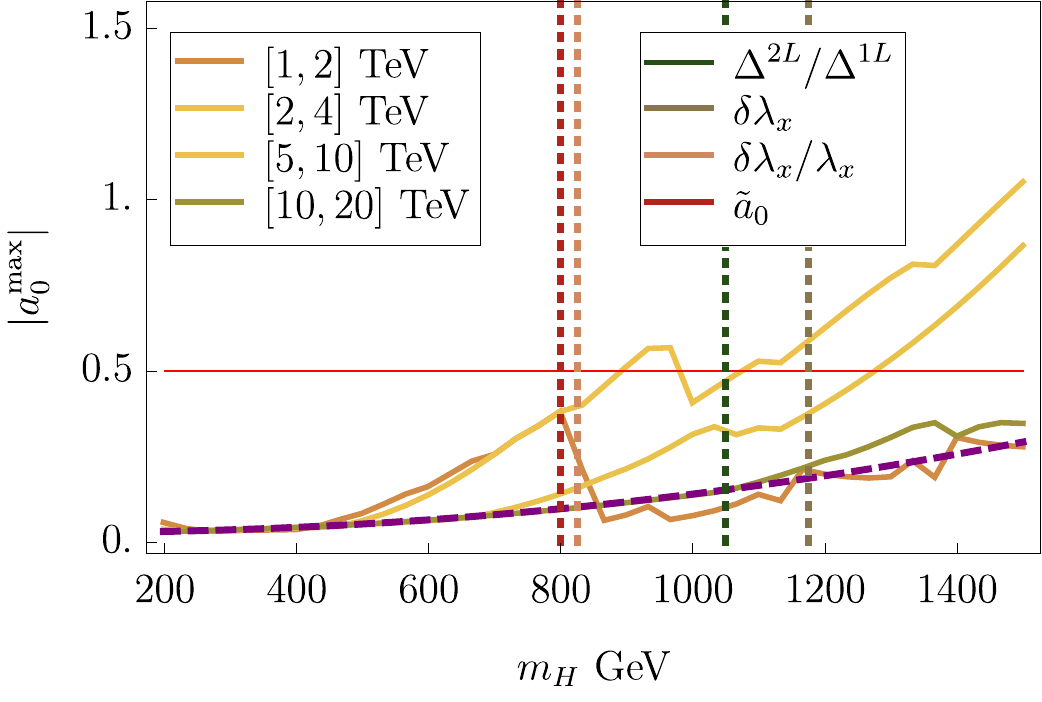} \hspace{.05\linewidth}
\includegraphics[width=.45\linewidth]{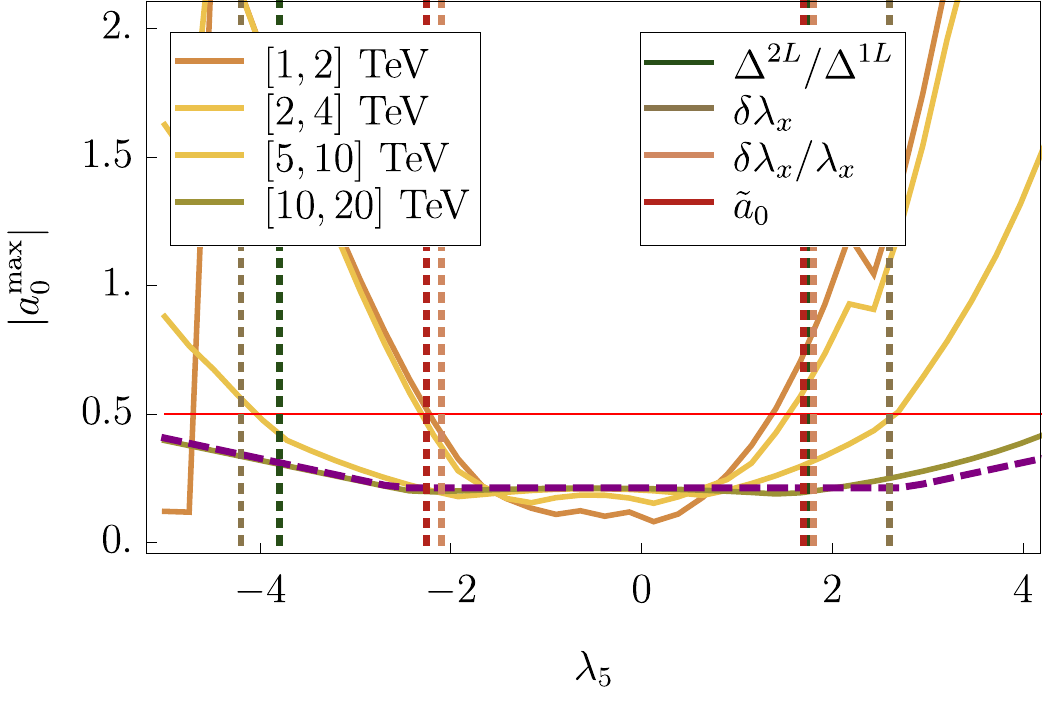} 
\caption{Comparison between the new unitarity constraints and the perturbativity constraints proposed in Ref.~\cite{Krauss:2017xpj}. The maximal eigenvalue of the scattering matrix calculated 
for different intervals of the scatter energy $\sqrt{s}$ are shown. The vertical lines show where the different perturbativity constraints, summarised in Eqs.~(\ref{eq:relative_size_CT}) to (\ref{eq:size_2loop_vs_1loop}), are violated. 
The purple dashed line gives the result using the old calculation with the large $s$ approximation. \\
{\emph{\it{Upper row:}}} $|a_0^{\rm max}|$ as a function of $\sin \theta_H$ for the input masses $m_H = 300\,$GeV (left-hand plot) and 800\,GeV (right-hand plot). The other parameters were set to $\lambda_2 = 0.1, \lambda_3=0.5,\lambda_4=-0.02,\lambda_5=0.1, \alpha=20\degree$.\\
{\emph{\it{Lower row:}}} $|a_0^{\rm max}|$ as a function of $m_H$ using $\lambda_2 = \lambda_3= \lambda_5=0.1,\lambda_4=-0.1, \alpha=20\degree, \sin\theta_H=0.25$ (left plane), and as a function of $\lambda_5$ using $\lambda_2 = 0.1, \lambda_3=0.5, \lambda_4=-0.1, \alpha=20\degree, \sin\theta_H=0.3, m_H=750$ (right plane).
 }
\label{fig:perturbativity_comparison}
\end{figure*}
We make this comparison for different parameter ranges which were discussed in Ref.~\cite{Krauss:2017xpj}, cf. Figs.~3, 6 and 7 in this reference. The results
are shown in Fig.~\ref{fig:perturbativity_comparison}. In the upper row, we vary $\sin\theta_H$ for two different heavy Higgs masses  $m_H=300,\,800\,$GeV. In the lower row we show the dependence on both $m_H$ and $\lambda_5$. For a 
discussion of the choices of varying parameters we refer the interested reader to Ref.~\cite{Krauss:2017xpj}.
All other input parameters are listed in the caption of each figure. We show the maximal scattering eigenvalue for different intervals of the scattering energy $\sqrt{s}$. 
All parameter regions which we show here are allowed by Higgs data, vacuum stability and the old unitarity constraints. However, 
in Ref.~\cite{Krauss:2017xpj} it was found that perturbation theory is not trustworthy in some ranges of the varied parameters. The large loop corrections are also caused by large (effective) trilinear couplings. Since the 
same couplings can also enhance the scattering amplitudes, we find that there is actually a nice agreement between the perturbativity constraints and the improved unitarity constraints discussed here. The reason for the very different scaling of the old and new unitarity constraints as a function of $\lambda_5$ are diagrams with effective trilinear couplings $\sim \lambda_5 v_{\rm SM}$. One finds that the dominant contributions scale as
\begin{equation}
|a_0^{\rm max}| \sim -\frac{2 \lambda_5^2 v_{\rm SM}^2 \log\left(\frac{m_H^2}{-3 m_H^2 + s}\right)}{3 \pi \sqrt{s (-4 m_H^2 + s)} s_H^2} \simeq \frac{1}{4\pi} \left(\frac{\lambda_5 v_{\rm SM}}{m_H s_H}\right)^2  < \frac12
\end{equation}
while the old constraints impose $\frac{\lambda_5}{8\pi}<\frac12$, ie. there can be a strong enhancement of $\lambda_5 \frac{v_{\rm SM}^2}{m^2_H s^2_H}$ for small $s_H$ and not too large $m_H$.

%

\subsubsection{RGE effects}
\begin{figure}[tb]
\includegraphics[width=\linewidth]{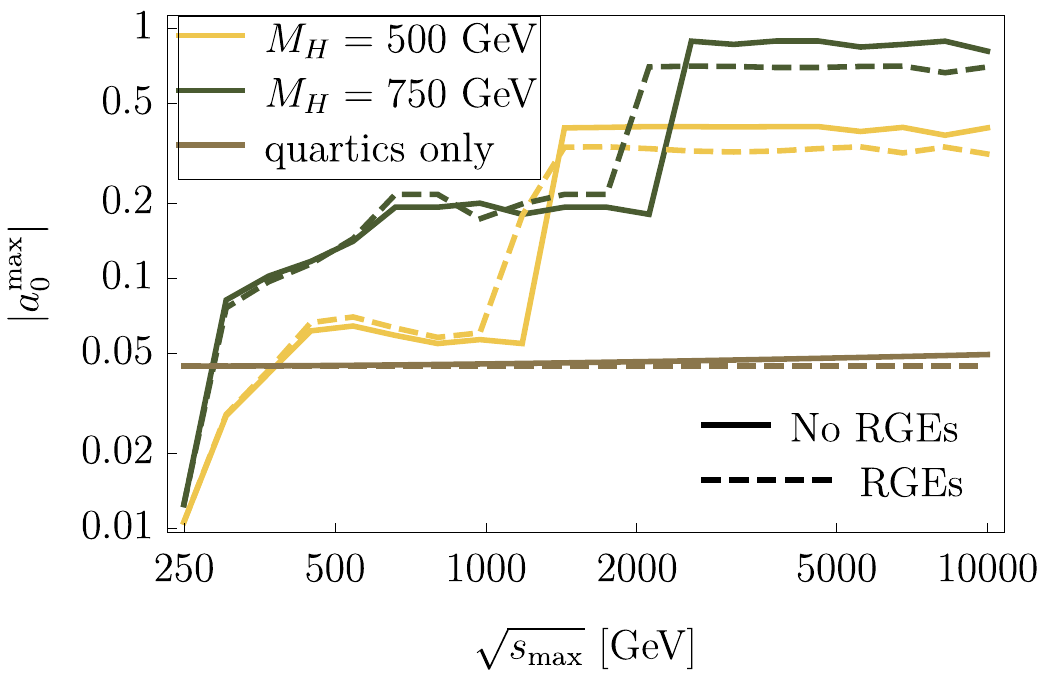} \\
\includegraphics[width=\linewidth]{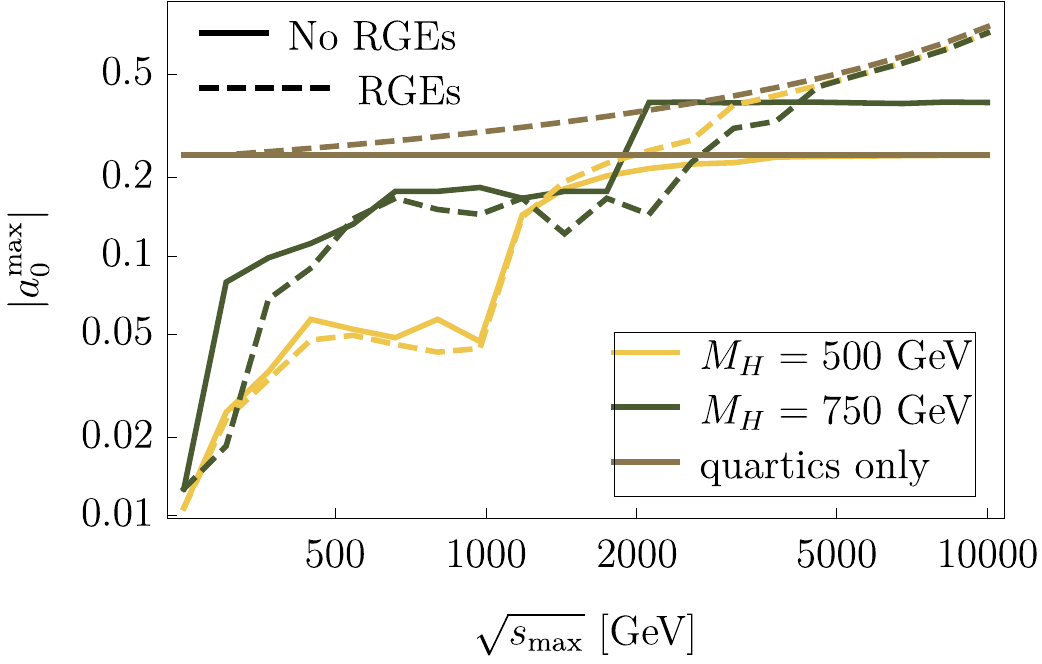} 
\caption{The maximal eigenvalue of the scattering matrix depending  on the highest scattering energy $\sqrt{s_{\rm max}}$ up to which we have computed the processes. We present the results with (dashed line) and without  (solid line) the inclusion of RGE running of the parameters.
In the first row we use small quartic couplings of $\pm0.1$, while in the second row the couplings are $\pm1$.  }
\label{fig:GM_RGEs}
\end{figure}

Finally, we want to estimate the effect of higher-order corrections which we are still missing by performing a pure tree-level calculation. A full one-loop calculation of the entire $2\to 2$ scattering process is well beyond the scope of this paper. Therefore, we use the one-loop RGEs and calculate the process with running couplings and masses at the scattering energy $\sqrt{s}$. The RGEs for the GM model have been calculated in Ref.~\cite{Blasi:2017xmc} using the generalised version of the Lagrangian parameters. We have implemented this model in \SARAH and cross-checked the $\beta$-functions. While we agree with the expressions for all scalar parameters, we found discrepancies for $g_2$ and the Yukawa couplings. We find for the $\beta$ functions of these parameters in the limit of $Y_\tau\to0$ and no flavour mixing 
\begin{align}
16 \pi \beta_{g_2} = & -\frac{13}{6} g_2^3 \,,\\
16 \pi \beta_{Y_t} = & \frac{9 Y_t^3}{2} + \frac{3 Y_b^2 Y_t}{2} -Y_t\left(\frac{17 g_1^2}{20}+\frac{9 g_2^2}{4}+8 g_3^2\right)\,,\\
16 \pi \beta_{Y_b} = &  \frac92 Y_b^3+\frac32 Y_t^2 Y_b -  Y_b \left(\frac14 g_1^2+ \frac94 g_2^2+8 g_3^2\right) \,.
\end{align}
Having the model and the RGEs at hand, we can check the impact of the running. Since we are here only interested in an estimate of the size of this effect, we use the simplified -- but common -- approach of tree-level matching combined with one-loop running. Higher order corrections will be important especially in the presence of large quartic couplings \cite{Braathen:2017jvs}. 
We show in Fig.~\ref{fig:GM_RGEs} the maximal eigenvalue of the scattering matrix as a function of the largest considered scattering energy $\sqrt{s_{\rm max}}$. For each $\sqrt{s_{\rm max}}$, we
check for the best scattering energy between $s$ which is in between $s_{\rm min}$ and $s_{\rm max}$ including the RGE running of all couplings up to $\sqrt{s}$. For comparison we also show $a_0^{\max}$ without RGE running.\footnote{{Per construction, the slope of the obtained curve can never be negative if no RGEs are considered. Small variations are only due to discrete steps in the scanning of the scattering amplitudes up to $\sqrt{s_{\rm max}}$.}} 
We do so for the heavy Higgs masses $m_H=500~\text{GeV}$ and $750~\text{GeV}$. The other parameters were set to
\begin{eqnarray}
& \lambda_2 = 0, \lambda_3 = -\lambda_4 = \lambda_5 = x \,\, \text{with}\, \, x= \{0.1, 1\} & \nonumber \\ 
&\alpha=20\degree, \sin\theta_H = 0.22 \,.& \
\end{eqnarray}
For comparison, we include also the maximal eigenvalue using the old calculation which includes only the contributions from point interactions. For those, we also include the RGE running up to a scale $Q=\sqrt{s_{\rm max}}$. This might look a bit strange: these constraints use already the large $s$ limit since they neglect the phase space factor $\sqrt{1-m^2/s}$; however this is a common procedure in literature and shall only serve for illustration purposes here. 
We see that for small quartic couplings, the effect of the RGEs is moderately small. This is not surprising because also the running of the trilinear parameters $M_1$ and $M_2$ is proportional to the quartic couplings as well as gauge and Yukawa couplings squared. Therefore, even if we run the quartic couplings to $10$~TeV, the contributions from the trilinear couplings dominate the unitarity constraints. 
If we go to larger values of the quartics (lower row in Fig.~\ref{fig:GM_RGEs}), the running becomes stronger because of the $\lambda^3$ dependence of the respective $\beta$ functions. Thus, the scale dependence of the old constraints is quite strong and to some extent also much stronger than the one of the new constraints. 
%
At large $\sqrt{s_{\rm max}}$, when the quartic couplings have grown even more through RGE evolution, their point interactions also dominate over the $2\to 2$ scattering including propagators -- leading to an agreement 
between the old and new constraints independently of the chosen value of $m_H$.

\subsection{Impact on benchmark scenarios}
Before we conclude, we want to comment briefly on the impact of the new constraints on benchmark scenarios studied in literature. 
One widely used benchmark plane is the so-called $(m_5,s_H)$ where by construction the trilinear couplings are small \cite{deFlorian:2016spz,Logan:2017jpr}. Therefore, the change 
in the scattering elements is only moderate. In contrast, very recently Ref.~\cite{Chiang:2018xpl} has proposed six benchmark points which cover also larger values of $M_1$ and $M_2$. Therefore, we find
quite significant changes. In particular BP2, BP4 and
BP6
 are clearly ruled out. For these three points, the maximal scattering eigenvalue changes by a factor 10 to 100. 

\section{Conclusion}
\label{sec:conclusion}

In this paper we have computed the full scalar $2\to 2$ scattering amplitudes for models with scalar $SU(2)_L$ triplets.
Those amplitudes are needed for determining the bounds from imposing perturbative unitarity. So far, these bounds have been computed using the limit of large scattering energy, neglecting all diagrams with internal propagators. Here, instead, we include the effects stemming from finite energies $\sqrt{s}$. 
We find that the full calculation is necessary in the presence of large trilinear scalar interactions, be it from electroweak symmetry breaking effects or from trilinear Lagrangian parameters. We showed this at the example of three models: (i) the SM extended by a real triplet with zero hypercharge, (ii) the SM extended by a complex triplet with $Y=1$, and (iii) the Georgi-Machacek model. In all examples we find sizeable regions of parameter space which are excluded by the constraints obtained from the full calculation but which would have been regarded as allowed using the old procedure. For the last model, we find good agreement of the new unitarity constraints with the recently proposed loop-level perturbativity checks of Ref.~\cite{Krauss:2017xpj}.
Since our study makes use of tree-level relations, we have finally estimated the effects of loop corrections by including renormalization group running of the parameters to the scattering energy. 

A more thorough estimate of the loop effects is beyond the scope of this paper and we leave the inclusion of higher-order corrections to the calculation of the scattering amplitudes, which can be important in particular in non-supersymmetric models, to future work.
The results for other very popular models like the singlet-extended SM and two-Higgs doublet models are discussed elsewhere \cite{Goodsell:2018tti,Goodsell:2018fex}.

\section*{Acknowledgements}
We thank Mark Goodsell for very fruitful discussions and collaboration on related work, and Sophie Williamson for collaboration in the early stages of this work.
FS is supported by ERC Recognition Award ERC-RA-0008 of the Helmholtz Association. MEK is supported by the DFG Research Unit 2239 ``New Physics at the LHC''. We further thank the LPTHE in Paris for their
hospitality while this work triggered.

\bibliography{lit}

\end{document}